\documentclass[pra,twocolumn,showpacs,amsmath,amssymb]{revtex4-1}
\usepackage{psfrag,graphicx}
\usepackage{amsfonts,amssymb,amsmath}      
\usepackage{graphicx}
\usepackage{dcolumn}
\usepackage{esvect}
\usepackage{bm}
\usepackage{hyperref}
\usepackage{color}
\usepackage{epstopdf}

\begin{document}


\title{Quantum enhanced spectroscopy with entangled multi-photon states}

\author{Hossein T. Dinani$^{1}$, Manish K. Gupta$^2$, Jonathan P. Dowling$^2$, Dominic W. Berry$^1$}

\affiliation{$^1$Department of Physics and Astronomy, Macquarie University, Sydney, NSW 2109, Australia
\\
$^2$Hearne Institute for Theoretical Physics, Department of Physics and Astronomy, Louisiana State University, Baton Rouge, Louisiana 70803, USA}

\date{\today}

\begin{abstract}
Traditionally, spectroscopy is performed by examining the position of absorption lines.  However, at frequencies near the transition frequency, additional information can be obtained from the phase shift.  In this work we consider the information about the transition frequency obtained from both the absorption and the phase shift, as quantified by the Fisher information in an interferometric measurement.  We examine the use of multiple single-photon states, NOON states, and numerically optimized states that are entangled and have multiple photons. We find the optimized states that improve over the standard quantum limit set by independent single photons for some atom number densities.
\end{abstract}

\maketitle
\section{introduction}

The goal of quantum metrology is to obtain the most precise measurements possible with minimal resources~\cite{Giovannetti}. Many types of high-precision measurement use a form of interferometry. In interferometers, the unknown parameter is imprinted as the relative phase between a superposition of states. Measurement of the output state, after quantum interference, gives information about the unknown parameter. \par

One particular example is optical interferometry with Mach-Zehnder interferometers. In this case, using $N$ independent photons gives $1/\sqrt{N}$ scaling for the uncertainty of phase measurements, which is known as the standard quantum limit. However, using $N$ entangled photons gives $1/N$ scaling for the phase uncertainty, which is often called the Heisenberg limit. This enhancement in sensitivity is of much importance in probing delicate systems such as atoms \cite{Wolfgramm} and biological samples \cite{{Taylor}}.\par


A well-known type of entangled states is NOON states \cite{{Sanders},{Dowling}}. NOON states saturate the Heisenberg limit. However, in the presence of absorption, NOON states perform poorly \cite{{Gilbert},{Rubin}} and other states must be considered \cite{{Dorner},{Dob},{Kac},{HTD}}.
Even with such states, the advantage over the standard quantum limit is reduced by loss. 
However, we can take advantage of the sensitivity of nonclassical properties of quantum states to absorption. The sensitivity of quantum coherence can be used efficiently to estimate absorption \cite{Scheel}, and also estimate physical quantities that the absorption depends on.

In Ref.~\cite{Whittaker} a sub-shot-noise measurement of absorption is obtained using heralded single photons. In that work, a non-interferometric setup was used, where all the information is obtained from absorption, and the quantum enhancement results from sub-Poissonian statistics of single photons. According to the Kramers-Kronig relation, absorption is accompanied by a phase shift \cite{Jackson}. However, the information from the phase is only accessible if we take advantage of superposition and interference.\par  

One would think, since NOON states have coherence which is highly sensitive to loss, they could be the best candidates to sense loss or a parameter which causes loss in the system. However, here we show that for the task of spectroscopy, when there is an advantage to using entangled states, it is possible to find more general entangled states that perform better than NOON states.\par

The system we are considering here is an ensemble of  atoms. We are interested in measuring a transition frequency of the atoms. If this ensemble is probed by a beam of photons, the absorption of photons, and phase shift imposed on the probe, both depend on the transition frequency of atoms. We consider a Mach-Zehnder interferometer with the atomic ensemble placed in one of the arms of the interferometer. We optimize over the state in the arms of the interferometer and find the state from which we obtain the maximum information about the atomic transition frequency.

\section{Interferometric scheme}\label{section-scheme}
\begin{figure}[t]
\centering
\includegraphics[scale=1.5]{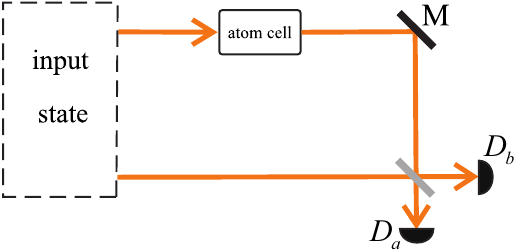}
\caption{A Mach-Zehnder interferometer with an ensemble of atoms placed in the upper arm. ${D}_a$ and ${D}_b$ are photon number detectors in the output modes.}
\label{fig-intf}
\end{figure}

Consider a Mach-Zehnder interferometer, as shown in Fig.~\ref{fig-intf}, with an ensemble of atoms placed in the upper arm of the interferometer. 
 Here, we consider an ensemble of identical two-level atoms in the absence of Doppler broadening and dipole dephasing. This simple model gives a good qualitative description of the problem. Assuming that all atoms interact equally with the input quantum state and that there is no interaction between atoms, using the dipole and rotating-wave approximation the susceptibility of the ensemble is given by \cite{{Kok},{Beausoleil}}
\begin{equation}\label{chi_realimag}
\chi (\Delta )={\chi }'(\Delta )+i{\chi }''(\Delta )  =   \frac{{2 \mathcal{N}{\mu ^2}}}{{\hbar {\varepsilon _0}}}\frac{{\Delta  + i\gamma_s }}{{{\Delta ^2} + {\gamma^2_s}}},
\end{equation}
where $\Delta=\omega-\omega_0$ is the detuning between $\omega_0$, the transition frequency of atoms, and $\omega$, the frequency of input photons, $\gamma_s$ is the spontaneous decay rate of the excited state, $\mathcal{N}$ is the number density of atoms, $\mu$ is the electric dipole moment, $\hbar$ is the reduced Planck constant and $\varepsilon_0$ is the vacuum permittivity. Details of the derivation of this susceptibility based on interaction of an ensemble of atoms with quantized light are given in Refs.~\cite{{Kok},{Beausoleil}}.\par
\begin{figure}[b]
\centering
\includegraphics[scale=0.9]{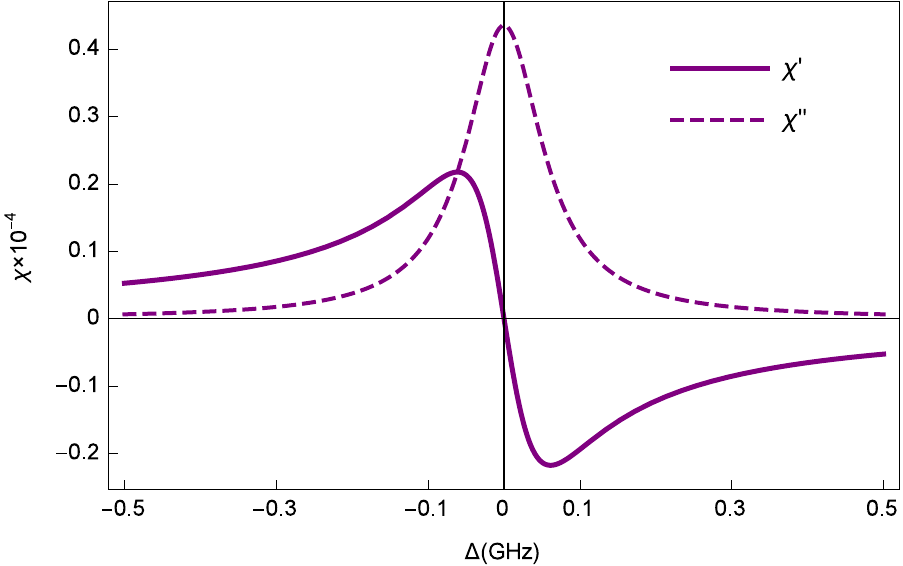}
\caption{Real (solid line) and imaginary (dashed line) parts of susceptibility, $\chi'$ and $\chi''$ respectively, for an ensemble of two level atoms.}
\label{fig-chi}
\end{figure}

 The imaginary and real parts of the susceptibility are plotted in Fig.~\ref{fig-chi}. In this figure we have used data for the D1 transition line of sodium from Ref.~\cite{Steck}; i.e.~$\mu=0.704\times10^{-29}$~${\rm C}\cdot{\rm m}$ and $\gamma_s=61.354\times 10^6$ s$^{-1}$. For the number density of atoms we have used $\mathcal{N}=2.5\times10^{16}$~m$^{-3}$.
\par
\begin{figure}[t!]
\centering
\includegraphics[scale=0.725]{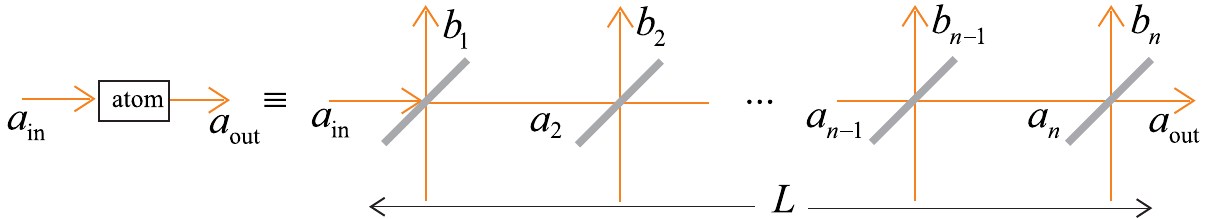}
\caption{Beam splitter model to model the interaction of photons with the ensemble of atoms.}
\label{fig-BSatom}
\end{figure}
Knowing the susceptibility of the atomic medium, the effect of the atomic ensemble in the upper arm of the Mach-Zehnder interferometer can be modeled by the beam splitter model proposed in Refs.~\cite{{Jeffers},{Loudon}}. Normally, one beam splitter is used to model loss in each of the arms of a Mach-Zehnder interferometer \cite{Dob, HTD, JD}. However, here we consider a line of $n$ beam splitters, shown in Fig.~\ref{fig-BSatom} where each beam splitter represents one of the atoms in the ensemble. The $k$th beam splitter transforms the creation operator $a^{\dag}_k$ according to
\begin{equation}
{{a}^\dag_{k}} = \sqrt{t(\omega )}{{a}^\dag_{k+1}}+\sqrt{r(\omega )}{{b}^\dag_{k}},
\end{equation}
where $t(\omega)$ and $r(\omega)$ are the transmissivity and reflectivity of the beam splitter, $\omega$ is the frequency of input photons, and $b_k$ is the loss mode of the $k$-th beam splitter.\par
The effect of the atomic ensemble is obtained by applying all the beam splitters, and taking the limit as the number of beam splitters approaches infinity.
The creation operator of the input mode $a^\dag_{\rm in}$ is therefore transformed to \cite{{Jeffers},{Loudon}}
\begin{eqnarray}\label{transformation}
a_{{\rm in}}^{\dagger } &=& a_{{\rm out}}^{\dagger }{{e}^{-i\frac{\omega L}{c}{\sqrt{1+\chi }}}} \nonumber \\
&&-i\sqrt{\frac{\omega }{c}{\chi }''}\int_{0}^{L}{{{e}^{-i\frac{\omega }{c}(L-z)\sqrt{{1+\chi }}}}{{b}^{\dagger }}(z)dz},
\end{eqnarray}
where $L$ is the length of the ensemble, $\omega$ is the frequency of the input photons, $c$ is the speed of light and $\chi''$ is the imaginary part of  the susceptibility of the atomic ensemble $\chi=\chi'+i\chi''$. The real part of susceptibility, $\chi'$, describes dispersion and the imaginary part, $\chi''$, describes absorption by the ensemble. \par
\begin{figure}[b!]
\centering
\includegraphics[scale=0.9]{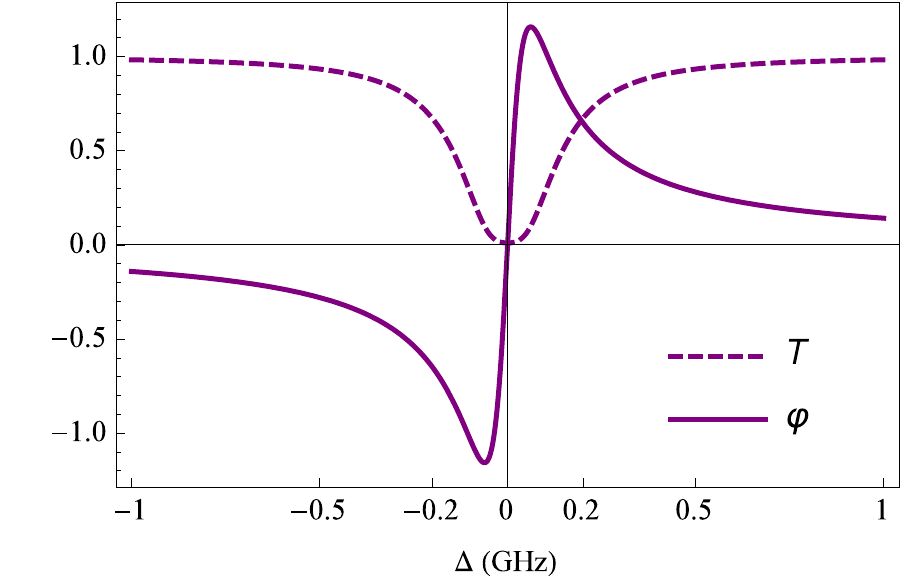}
\caption{Transmissivity $T$ (dashed line) and phase shift $\varphi$ (solid line) vs. detuning $\Delta$. }
\label{fig-tphi}
\end{figure}
From Eq.~\eqref{transformation}, the transmissivity of the ensemble, $T$, and the phase shift imposed on the state from the ensemble,  $\varphi$, can be written in terms of the imaginary and real parts of the susceptibility
\begin{equation}\label{Tphi}
T  = {e^{ - \chi ''\omega L/c}}, \qquad \varphi  =  - \frac{{\chi '\omega L}}{{2c}}.
\end{equation}
Here, we have used the approximation $\sqrt{1+\chi} \approx 1+ \chi/2$. The quantities $T$ and $\varphi$ are plotted in Fig.~\ref{fig-tphi}. This figure is plotted for $\omega_0=2 \pi (508.33)$~THz, which is the D1 transition line of sodium~\cite{Steck}, and $L=1$~cm. As can be seen from Fig.~\ref{fig-tphi}, for detunings close to zero, $\varphi$ has the highest slope. However,  in this region, $T$ is very small.\par
In the following section we find the optimal states to measure the transition frequency of the atoms i.e.,~$\Delta$ in this scheme.
\section{Optimized states}\label{section-measurement}
 We consider the general form of the state in the arms of the interferometer to be
\begin{equation}\label{inputstate}
\left| \psi  \right\rangle = \sum\limits_{k = 0}^N {{\psi_k}\left| {N - k,k} \right\rangle },
\end{equation}
i.e.,~a pure state with the total photon number of $N$. We use Fisher information as the measure to quantify the metrological value of the states. 
According to the Cram\'{e}r-Rao bound \cite{QCRB} the variance in estimating a parameter, $\Delta$ in this case, using an unbiased estimate, is lower bounded by the inverse of the Fisher information $F(\Delta)$
\begin{equation}
\operatorname{var}(\Delta )\ge 1/{F(\Delta )}.
\end{equation}
 Here, we are considering photon number detection in the output modes, thus we are using classical rather than quantum Fisher information. The Fisher information represents the amount of information about $\Delta$ contained in the measurement results. It is given as
\begin{equation}
F{\left( \Delta  \right)}=\sum\limits_{{{n}_{1}},{{n}_{2}}}^{{}}{\frac{1}{{{P}_{{{n}_{1}},{{n}_{2}}}}\left( \Delta  \right)}{{\left( \frac{\partial {{P}_{{{n}_{1}},{{n}_{2}}}}{\left( \Delta  \right)}}{\partial \Delta } \right)}^{2}}},
\end{equation}
where $P_{n_{1},n_{2}}(\Delta)$ is the probability of detecting $n_1$ and $n_2$ photons in each of the output ports.\par
\begin{widetext}
 Considering the state given in Eq.~\eqref{inputstate}, applying the atom cell transformation given in Eq.~\eqref{transformation} on the first mode, and the last 50/50 beam splitter of the interferometer on both modes, we obtain 
\begin{eqnarray}\label{prob}
 && {{P}_{{{n}_{1}},{{n}_{2}}}}{\left( {\omega}  \right)}=\sum\limits_{k=0}^{{{n}_{1}}+{{n}_{2}}}{\sum\limits_{{k}'=0}^{{{n}_{1}}+{{n}_{2}}}{\sum\limits_{u={{n}_{2}}-k}^{{{n}_{2}}}{\sum\limits_{{{v}_{}}={{n}_{2}}-{k}'}^{{{n}_{2}}}{{{\psi }_{k}}\psi _{{{k}'}}^{*}\frac{{{n}_{1}}!{{n}_{2}}!(N-{{n}_{1}}-{{n}_{2}})!}{\sqrt{k!{k}'!(N-k)!(N-{k}')!}}{{\left( \frac{1}{2} \right)}^{{{n}_{1}}+{{n}_{2}}}}(-1)^{k-n_2}}}}}  \nonumber \\
 &&\qquad \qquad\qquad \times {\scriptscriptstyle \left( \begin{matrix}
   N-{k}'  \\
   N-{{n}_{1}}-{{n}_{2}}  \\
\end{matrix} \right)\left( \begin{matrix}
   N-k  \\
   N-{{n}_{1}}-{{n}_{2}}  \\
\end{matrix} \right)\left( \begin{matrix}
   {{n}_{1}}+{{n}_{2}}-k  \\
   u  \\
\end{matrix} \right)\left( \begin{matrix}
   {{n}_{1}}+{{n}_{2}}-{k}'  \\
   v  \\
\end{matrix} \right)\left( \begin{matrix}
   k  \\
   k+u-{{n}_{2}}  \\
\end{matrix} \right)\left( \begin{matrix}
   {{k}'}  \\
   {k}'+v-{{n}_{2}}  \\
\end{matrix} \right) } \nonumber\\ 
 && \qquad \qquad \qquad \times {{(1-e^{-\omega L \chi''/c})}^{N-{{n}_{1}}-{{n}_{2}}}}{{e}^{i\omega {\chi}' L(k-{k}')/(2c)}}{{e}^{-\omega L{\chi }''(2{{n}_{1}}+2{{n}_{2}}-k-{k}')/(2c)}},
\end{eqnarray}
\end{widetext}
where $\omega=\Delta+\omega_0$ is the frequency of the photons in the input state. Because we are quantifying the metrological value of the states via the Fisher information, we regard the optimal states to be those which maximize the Fisher information. We have found the optimal values of $\psi_k$ numerically using the particle swarm optimization (PSO) algorithm.\par

In the PSO algorithm a swarm of particles search the space of $\psi_k$ coefficients for those that maximize the Fisher information. Each particle has a velocity $\vv{v}_{}$ and position $\vv{x}_{}$ which are updated to $\vv{v}'$ and $\vv{x}'$ according to its best previous position $\vv{x}_{\ell}$, and the best position of the entire swarm $\vv{x}_{g}$ as \cite{PSO}
\begin{eqnarray}\label{eq-PSO}
&&\qquad \qquad \qquad {\vv{x}'_{}}= {\vv{x}_{}} + {\vv{v}'_{}}, \nonumber \\
&&{\vv{v}'_{}}= \chi \left[ {{\vv{v}_{}} + {c_g}{r_g}\left( {{\vv{{x}_{g}}} - {\vv{x}_{}}} \right) + {c_{\ell}}{r_{\ell}}\left( {{\vv{{x}_{\ell}}} - {\vv{x}_{}}} \right)} \right].
\end{eqnarray}
Here, $r_{g}$ and $r_{\ell}$ are uniform random numbers in the interval $[0,1]$, and $\chi$, $c_g$ and $c_{\ell}$ are constants. In our simulations, we used $\chi=0.729$, $c_{\ell}=c_g=2.05$ with 10 particles and 100 iterations.\par

\begin{figure*}[t!]
\centering
\includegraphics[scale=0.72]{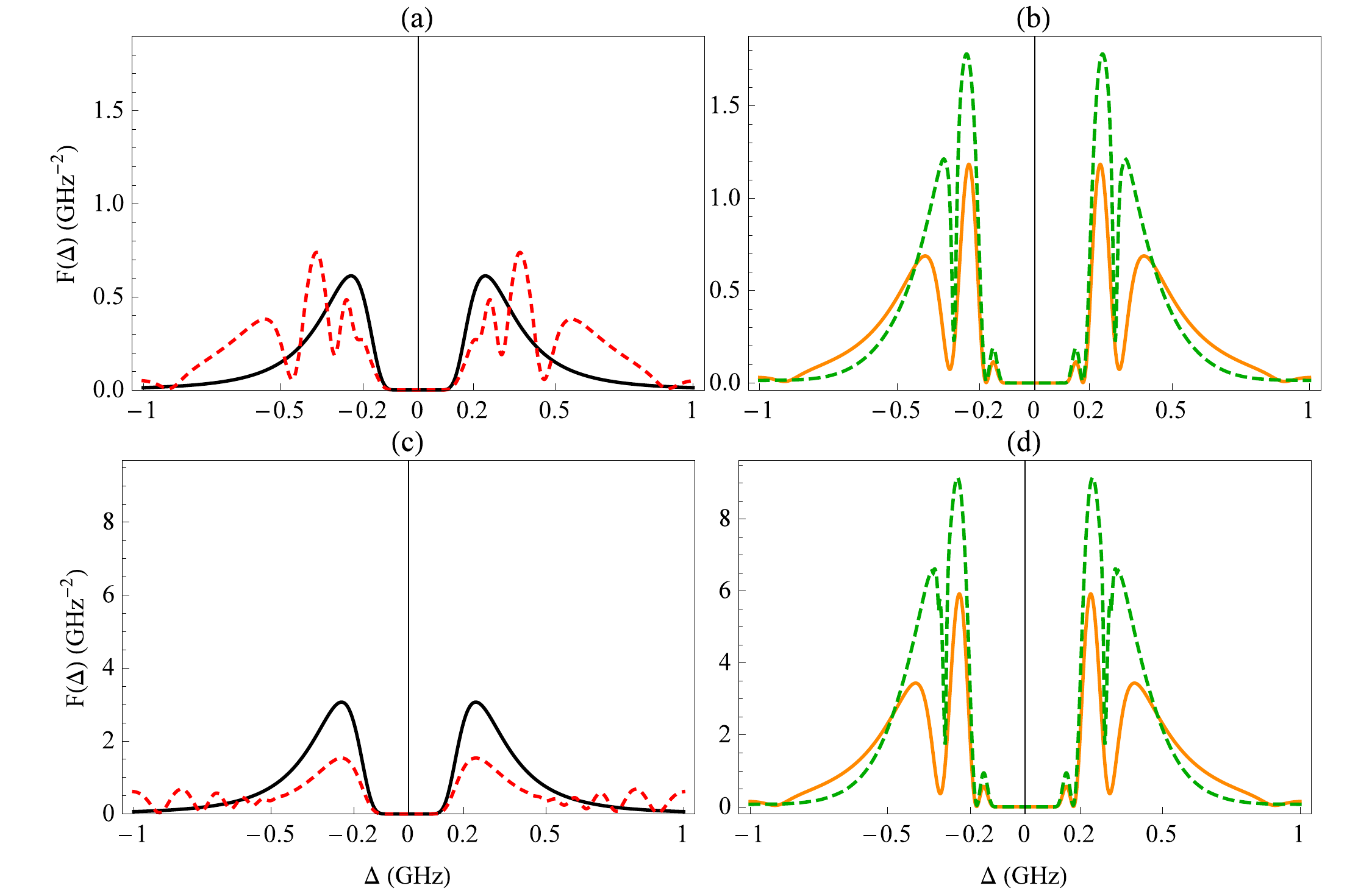}
\caption{Fisher information $F(\Delta)$ versus detuning $\Delta$ for $N=2$ photons (upper row) and $N=10$ (lower row). Solid-black line: $N$ independent single photons ${\left| {1,0} \right\rangle ^{ \otimes N}}$. Dashed-red line: $N$-photon NOON state  $\left( {\left| {N,0} \right\rangle  + \left| {0,N} \right\rangle } \right)/\sqrt 2$. Dashed green line: $N$-photon optimal state. Solid-orange line: $N$ copies of single-photon NOON states $\left( {\left| {1,0} \right\rangle  + \left| {0,1} \right\rangle } \right)/\sqrt 2 $. }
\label{fig-FIN210}
\end{figure*}
We have found that the optimal state for a specific type of atoms, only depends on the product of the number density of atoms and cell length, $\mathcal{N}L$. This can be explained in the following way. In Eqs.~\eqref{Tphi} and \eqref{prob}, we have $\omega \chi' L$ and $\omega \chi'' L$ which can be written as
\begin{equation}
\omega \chi L  = \omega L (\chi'+\chi'') =\frac{2\mathcal{N}L{{\mu }^{2}}{{\omega }_{0}}}{\hbar {{\varepsilon }_{0}}{{\gamma }_{s}}}\frac{\left( \Delta /{{\gamma }_{s}}+i \right)\left( 1+\Delta /{{\omega }_{0}} \right)}{1+{{\left( \Delta /{{\gamma }_{s}} \right)}^{2}}}.
\end{equation}
 For a given type of atom the multiplying factor at the front can only be varied via ${\mathcal{N}}$ or $L$. The other parameters, $\mu$, $\omega_0$ and $\gamma_s$ can be varied by changing the type of atom. These parameters affect the variation of $\omega\chi$ in three ways:
\begin{enumerate}
\item They change the multiplicative factor at the front.  As that factor can also be changed by varying ${{\cal N}}$ or $L$, that does not give any qualitatively different results than simply changing ${{\cal N}}$ or $L$.
\item The parameter $\gamma_s$ appears in the ratio $\Delta/\gamma_s$, and therefore provides a scaling to the variation of $\omega\chi L$ with $\Delta$. It therefore does not qualitatively change the results.
\item The parameter $\omega_0$ appears in the factor $(1+\Delta/\omega_0)$.  This factor affects the variation very little, because we consider a parameter regime where $\Delta/\omega_0 \ll 1$.
\end{enumerate} 

 In the following we keep $L$ constant at $1$~cm and discuss the two cases: $\mathcal{N}>10^{17}$~m$^{-3}$ (large $\mathcal{N}$) and  $\mathcal{N}<10^{17}$~m$^{-3}$ (small $\mathcal{N}$).

\subsection{Large $\mathcal{N}$}
For $\mathcal{N}>10^{17}$~m$^{-3}$, we have found that numerically optimized states of the form given in Eq.~\eqref{inputstate} perform better than NOON states and independent single photons.  
In Fig.~\ref{fig-FIN210} we have compared the Fisher information of the $N$-photon optimal state, $N$ independent single photon states ${\left| {1,0} \right\rangle}^{\otimes N}$, $N$ copies of a single-photon NOON state $\left( {\left| {1,0} \right\rangle  + \left| {0,1} \right\rangle } \right)/\sqrt 2 $, and an $N$-photon NOON state ${\left( {\left| {N,0} \right\rangle}  + {\left| {0,N} \right\rangle } \right)}/\sqrt 2$. This figure is plotted for $N=2$ (upper row) and $N=10$ (lower row).  In this figure, we have used $\omega_0=2\pi(508.332)$~THz, which is the transition frequency of the D1 line of sodium \cite{Steck}, and an atom density of $\mathcal{N}=2.5\times10^{17}$~m$^{-3}$.\par
Figure~\ref{fig-FIN210} shows that, even for $N=2$, the enhancement obtained by optimal states is significant. For larger photon numbers, as is shown in the graphs for $N=10$, there is no further significant improvement in the enhancement of the optimal states. Moreover, the optimal states with high photon numbers are not experimentally achievable with the current technology. On the other hand, it may be possible to generate the optimal states for $N=2$ with a  scheme similar to the one proposed in Ref.~\cite{HTD}. \par
Note that, close to resonance, for copies of single-photon NOON states the maximum peak is higher than for independent single photons and $N$-photon NOON states. This is as would be expected, since single-photon NOON states are the least sensitive NOON states to loss.
 From Fig.~\ref{fig-FIN210}(c), we see that ten-photon NOON states perform worse than independent single photons close to resonance. However, far from resonance, their Fisher information is even higher than the numerically obtained optimal states.  The reason why this is possible is that the optimal states are only optimal in the sense of giving the largest maximum Fisher information, but it is possible for other states to have larger Fisher information for detunings where the optimal states do not give their maximum Fisher information. On the other hand, as can be seen in Fig.~\ref{fig-FIN210}(a), two-photon NOON states are less sensitive to loss (compared to ten-photon NOON states), and close to resonance they perform better than independent single photons.\par

The other thing to note from Fig.~\ref{fig-FIN210} is that to be able to work in the region with maximum Fisher information  we need to have prior knowledge of the detuning. This is because the peaks of maximum Fisher information are quite narrow. In other words this scheme could be used to measure hyperfine splitting of atomic levels, or measure external effects, such as magnetic field,  on the transition frequency of atoms.\par
\begin{figure}[t!]
\centering
\includegraphics[scale=1]{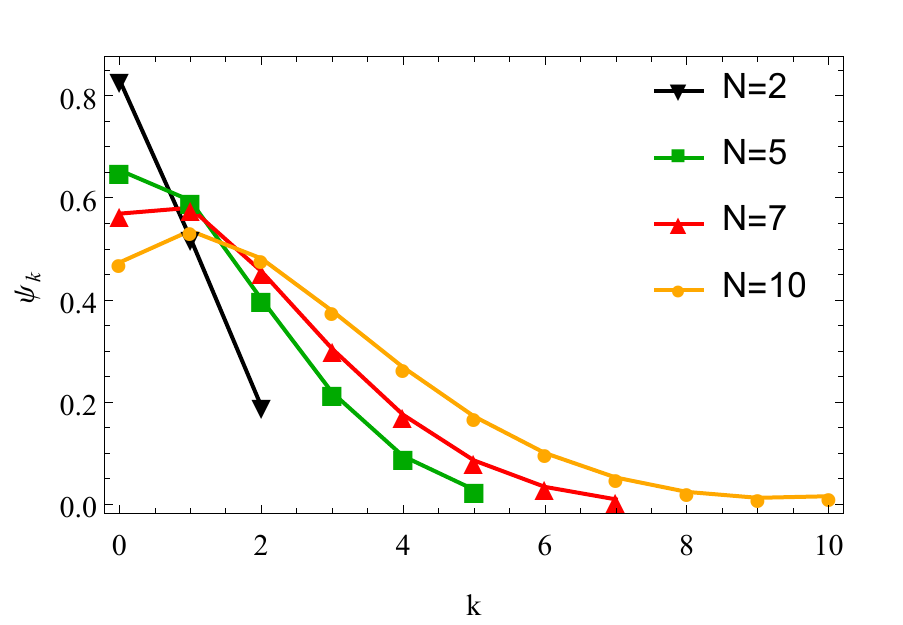}
\caption{Coefficients $\psi_k$ of the optimal states, for four values of the total photon number $N$.}
\label{fig-amp}
\end{figure}

In Fig.~\ref{fig-amp} we have plotted the values of the coefficients of the optimal states, $\psi_k$ in the superposition \eqref{inputstate}, for a range of photon numbers from $N=2$ to $N=10$. This figure shows that the optimal states have higher amplitudes for the terms with higher photon numbers in the arm that contains the atomic ensemble. That is, when there are more photons in the arm with the ensemble, they are more likely to be lost, giving more information about $\Delta$.\par

\subsection{Small $\mathcal{N}$}
\begin{figure}[h!]
\centering
\includegraphics[scale=0.5]{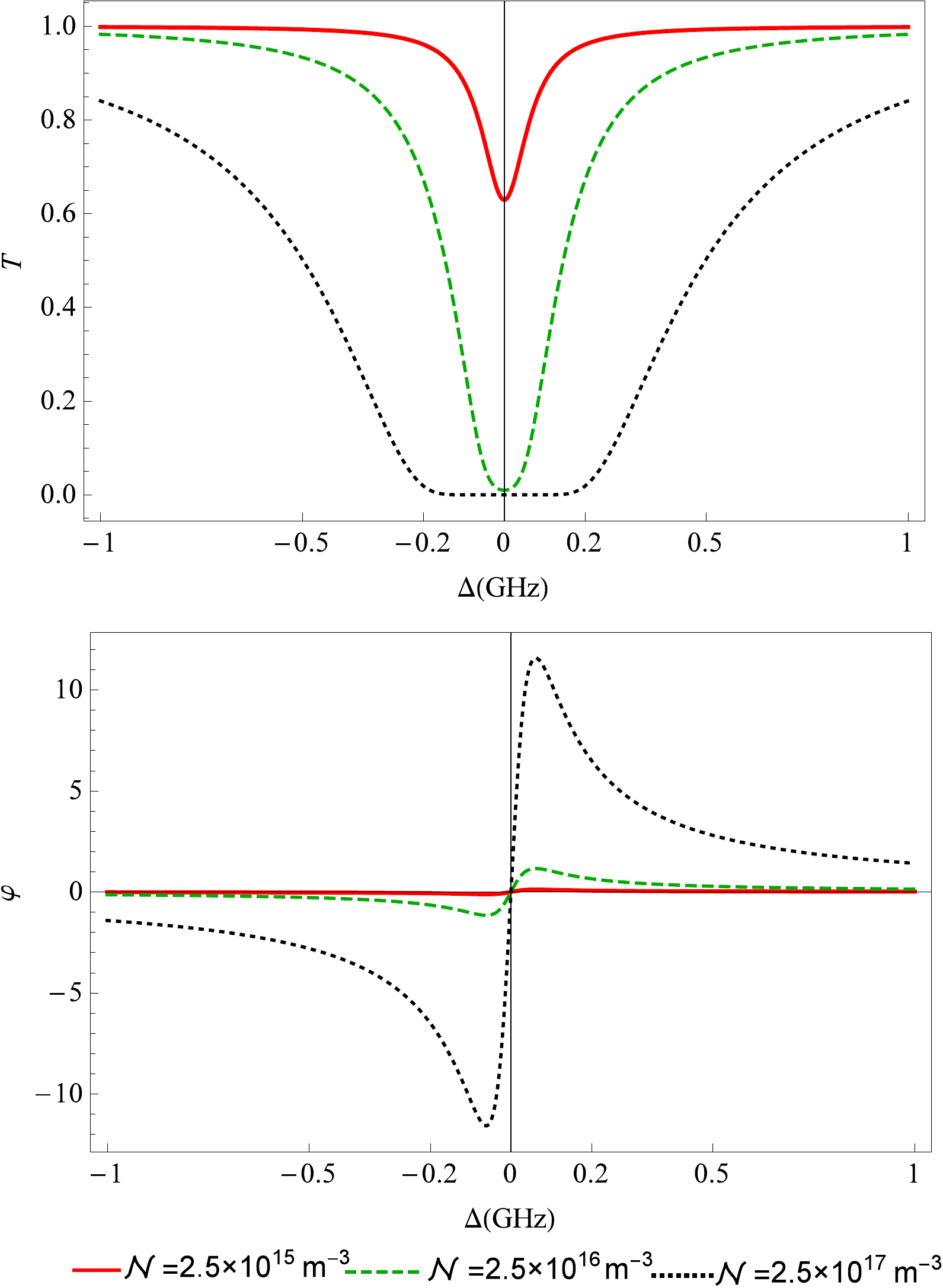}
\caption{Transmissivity $T$ and phase shift $\varphi$ versus $\Delta$, for a range of values of number density of atoms $\mathcal{N}$.}
\label{fig-J1}
\end{figure}
\begin{figure}[t!]
\centering
\includegraphics[scale=1]{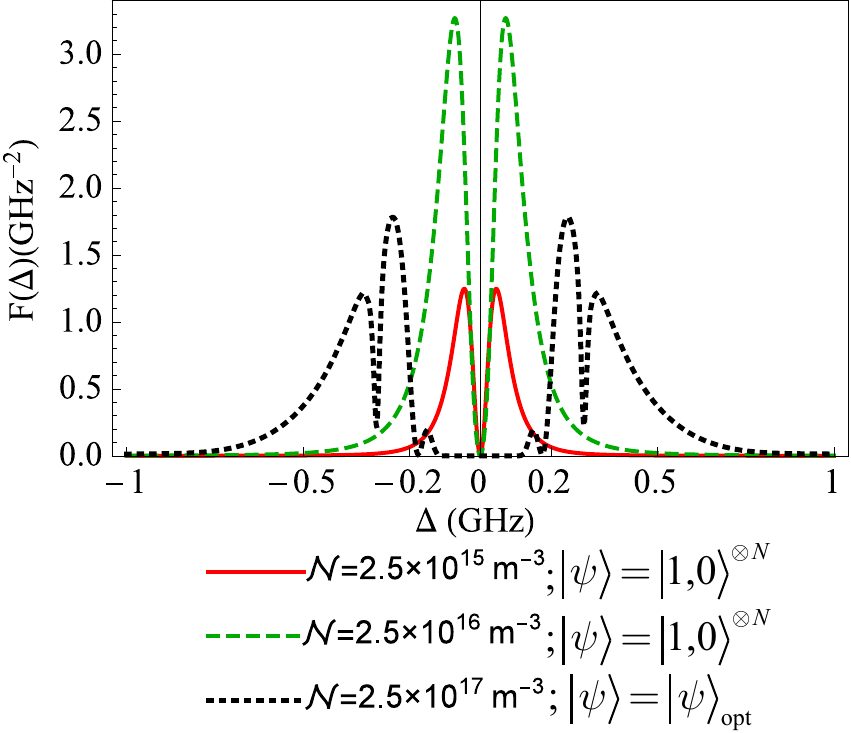}
\caption{Fisher information versus $\Delta$ for total number of photons $N=2$, for a range of number densities of atoms $\mathcal{N}$. Solid, red line:  $N$ independent single photons with $\mathcal{N}=2.5\times 10^{15}$~m$^{-3}$. Dashed, green line: $N$ independent single photons with $\mathcal{N}=2.5\times 10^{16}$~m$^{-3}$. Dashed, black line: numerically optimized state with $\mathcal{N}=2.5\times 10^{17}$~m$^{-3}$. }
\label{fig-FIs}
\end{figure}
For smaller values of $\mathcal{N}$ than considered in the previous subsection, the range of the phase shift is smaller (see Fig.~\ref{fig-J1}).
 In this case, the optimal state is $N$ independent single photons, ${\left| {1,0} \right\rangle}^{\otimes N} $. Having all the photons in the upper arm, only the loss is being probed, and no information is being obtained from the phase shift. The phase shift must be significant so that we can take advantage of interferometric schemes in spectroscopy. Surprisingly for $\mathcal{N}= 2.5 \times 10^{16}$ m$^{-3}$ the maximum of the Fisher information for $N$ independent single photons is even higher than the maximum of the Fisher information for the $N$-photon numerically optimized states with a larger number density of atoms which were considered in the previous subsection (see Fig.~\ref{fig-FIs}).\par

This could be understood from the variation of the transmissivity $T$ and phase shift $\varphi$ with $\mathcal{N}$, shown in Fig.~\ref{fig-J1}. For smaller values of $\mathcal{N}$ the range of the phase shift is also smaller, which eliminates the advantage in using entangled states. In this case, the Fisher information is coming from the variation in the absorption. As $\mathcal{N}$ is decreased further, the dip in the absorption is reduced which results in a smaller Fisher information. For the higher densities, there is a larger phase shift, but it is in a region where the absorption is very high.

\section{Conclusion}
In this work we found optimal multi-photon states for measurement of the transition frequency of atoms. The scheme proposed here is an interferometric scheme with photon number detection in the output. In order to find the best states for measurement of the transition frequency, we numerically optimised for the states that provide the largest Fisher information.\par

For the number density of atoms we considered initially, the imposed phase on the probe is large, and it is advantageous to using information from both the absorption and the phase shift for measuring the transition frequency. In this case, the optimal state is an entangled multi-photon state. This optimal state has a large weighting on the state with all photons in the arm with the atomic ensemble. On the other hand, for a smaller number density of atoms, the phase shift imposed on the probe is small and therefore the information from the phase shift is not significant enough to give any advantage. In this case, it is advantageous to pass all the photons through the atom cell and obtain all the information from absorption. 

Surprisingly there is a value of the number density, $\mathcal{N}=2.5 \times 10^{16}$~m$^{-3}$, for which $N$ independent single photons have the highest Fisher information; even higher than the Fisher information for the $N$-photon numerically optimised states with a larger number density of atoms.

\par
\section{Acknowledgements}
MKG and JPD would like to acknowledge funding from the Air Force Office of Scientific Research, The Army Research office, The National Science Foundation, and the Northrop Grumman Corporation. DWB is funded by an ARC Future Fellowship (FT100100761) and a Discovery Project (DP160102426).

\end{document}